\documentclass[aps,prl,preprint,superscriptaddress,
showpacs,showkeys]{revtex4-1}

\usepackage{graphicx} \usepackage{url} \usepackage{amsmath}
\usepackage{color}
\usepackage{subcaption}
\usepackage{multirow}
\usepackage{url}

\begin{document}
\preprint{SLAC-PUB-17483}
\title{Machine learning for design optimization of storage ring nonlinear dynamics}

\author{Faya Wang}
\affiliation{SLAC National Accelerator Laboratory, Menlo Park, CA 94025, USA}
\author{Minghao Song}
\affiliation{Illinois Institute of Technology, Chicago, IL 60616, USA}
\affiliation{SLAC National Accelerator Laboratory, Menlo Park, CA 94025, USA}
\author{Auralee Edelen}
\author{Xiaobiao Huang}
\email[]{xiahuang@slac.stanford.edu} 
\affiliation{SLAC National Accelerator Laboratory, Menlo Park, CA 94025, USA}

\date{\today}

\begin{abstract}
A novel approach to expedite design optimization  of nonlinear beam dynamics in storage rings is proposed and demonstrated in this 
study. At each iteration, a neural network surrogate model is
used to suggest new trial solutions in a multi-objective optimization task. 
The surrogate model is then updated with the new solutions, and this process is repeated until the final optimized solution is obtained.
We apply this approach to optimize the nonlinear beam dynamics of the SPEAR3 storage ring, where sextupole knobs are adjusted to simultaneously improve the dynamic aperture and the momentum 
aperture. The approach is shown to converge to the Pareto front considerably faster than the genetic and particle swarm
algorithms. 
\end{abstract}


\maketitle

The design of a complex system often requires searching for a globally optimal solution in a 
multi-dimensional parameter space. The dependence of the performance of the system on the parameters 
may be nonlinear, and the effects of the parameters on the performance may be 
coupled. Several measures of performance may need to be optimized simultaneously. 
For a system that involves complicated physical processes, the performance of a design solution is 
often evaluated through detailed physics modeling and simulation. The evaluation of a solution 
can be computationally expensive, e.g. taking hours on a large-scale cluster or high-performance computing system. 
Therefore, it is essential for the multi-objective optimization algorithms used in design optimization to 
have high efficiency. 

The optimization of nonlinear beam dynamics for storage rings is an excellent example that illustrates the 
need for high-efficiency optimization algorithms. The operation of a storage ring requires a large dynamic aperture (DA) and
a large local momentum aperture (LMA) throughout the ring. A large DA allows full capture of the beam injected into the storage ring, 
while a large momentum aperture leads to a long Touschek lifetime for high-charge, low-emittance beams. 
In a low-emittance storage ring, the DA and LMA are limited by nonlinear 
transverse beam motion. Sextupole magnets, which are needed in the storage ring to correct 
the energy dependence of the betatron tunes (i.e., chromaticity), are the major source of nonlinear forces that 
perturb the beam motion. Under these nonlinear perturbations, the particle motion at large oscillation 
amplitude  becomes unstable and eventually causes beam loss to the vacuum chamber. 
Proper placement of the sextupoles in the lattice can result in cancellation of
the nonlinear perturbations from certain sextupoles. 
An important aspect in the art of lattice design is to arrange the sextupole 
magnets in a way that minimizes the nonlinear perturbations. 

As new storage ring designs push for ultra low emittances, achieving acceptable DA and LMA becomes increasingly 
more challenging~\cite{APSU}. This is because a low emittance requires small dispersion, which in turn requires stronger 
sextupoles for chromaticity correction. Additional nonlinear magnets are often introduced to the lattice in 
order to gain extra control parameters (i.e., "knobs") for controlling the nonlinear resonances. Extra knobs can also be created by 
adding  power supplies to allow independent variations for smaller groups of magnets. 
Obtaining optimal solutions to simultaneously achieve large DA and LMA with the available nonlinear dynamics 
control knobs is a challenging multi-objective optimization problem. 
Multi-objective genetic algorithms (MOGA) are frequently used to optimize nonlinear lattice designs~\cite{BorlandGA,YangGANSLS2}, and 
a commonly used MOGA algorithm is NSGA-II~\cite{NSGA2}. 
Particle swarm optimization (PSO)~\cite{PSO985692Kennedy} has also been used for lattice design optimization~\cite{HuangPSO}. 

MOGA and PSO are powerful methods for finding globally optimal solutions in a 
high-dimensional parameter space. 
However, these stochastic algorithms are not very efficient, as the new trial solutions are generated through random operations on the existing solutions (e.g. without exploiting any modeled relationships between input and output parameters). Thousands or tens-of-thousands of solutions may need to be evaluated before 
the optimal solutions 
can be found. 
The evaluation of a lattice solution is typically done through multi-particle tracking simulation, with 
which the DA and LMA are determined. Because particles may approach the nonlinear resonances that limit 
DA and LMA long after injection or Touschek scattering, it is necessary to track many turns to be sure the surviving particles are truly stable (this is normally on the order of one 
damping time). 
Evaluation of DA and LMA for one lattice solution by particle tracking could take tens of minutes. 
Therefore, one round of nonlinear lattice optimization can take days or weeks, despite the fact MOGA and PSO 
can take advantage of parallel computing by evaluating the solutions in the same generation simultaneously. 
Improving the efficiency of multi-objective optimization algorithms would substantially benefit storage ring lattice 
design practice, as well as many other design optimization problems for particle accelerators. 

In this study we demonstrate a novel optimization method that involves training neural networks (NNs) to approximate the 
objective functions and searching for the globally optimal solutions with the NN model. 
Using NN-based surrogate models for design optimization of accelerators has previously been investigated ~\cite{EdelenML}, where a surrogate model trained on a uniform random sampling of the input parameter space is used as a substitute for the physics model. The surrogate model is subsequently used in multi-objective optimization of the system. In our approach, the training and optimization processes are repeated iteratively, and during each iteration the NN model is updated with newly-evaluated trial solutions. 
This approach should in principle be more efficient and produce higher-quality final solutions, because the later generations will be more concentrated around the region with good solutions. For complex objective functions on a high dimensional parameter space, the benefit could be significant. 




For an optimization problem with $P$ decision variables and $M$ objectives, the goal is to find the Pareto front, i.e., 
the set of solutions which are better than all other solutions in at least one objective but are no better than each other. 
Each solution is represented by a vector ${\bf X}=(x_1, x_2, \cdots, x_P)$, where $x_i$, $i=1$, $2$, $\cdots$, $P$, 
are the decision variables. For practical optimization, each decision variable has a valid range. We normalize 
all parameter ranges to $[0, 1]$. The performance of a solution is represented by the function values of the 
objectives, ${\bf Y}=(y_1, y_2, \cdots, y_M)$, with $y_i=f_i({\bf X})$, for $i=1$ through $M$. 
We consider a minimization problem and assume the special case with $M=2$. 

MOGA and PSO algorithms utilize a population of solutions to explore the parameter space. The initial population may be generated 
randomly. Subsequently new trial solutions are generated and evaluated based on the quality of the previous solutions. 
Non-dominated sorting~\cite{NSGA2} is applied to the combined set of the 
existing best solutions and the new solutions to update the population of 
best solutions. Since the new solutions are generated with the guidance of the known good solutions, 
their performance on average tends to improve over previous generations. The final optimal solution lies on a Pareto front that defines the optimal trade-off achievable between competing objectives (in our case, DA and LMA).

In our NN-based method, the evaluated solutions are used to train and update surrogate models of the functions to be optimized. New trial solutions are generated by optimization of the surrogate model. 
The algorithm may be summarized as follows:


\noindent
\textbf{Start}: Set generation index $g=0$. 
Randomly generate a population of $N$ solutions within the parameter space and
evaluate their performance with simulation
to obtain the initial data set,
${\mathcal D}_0=\{({\mathbf X}^0_i,{\mathbf Y}^0_i)|i=1, 2, \cdots N\}$. 
Train the initial NN model, ${\mathbf Y}={\mathcal M}_0({\mathbf X})$, using ${\mathbf X}^0$ as input and 
${\mathbf Y}^0$ as output.

\noindent
\textbf{Repeat}: 


1. Obtain a small set of solutions on the Pareto front for the multi-objective 
optimization problem, ${\mathcal M}_g$, using optimization methods for single-objective optimization. The multiple objective functions are 
combined into one objective with a weighted sum. For the case of $M=2$, the objective is 
\begin{eqnarray}
f_w = w\frac{f_1-f_{1a}}{f_{1b}-f_{1a}}+(1-w)\frac{f_2-f_{2a}}{f_{2b}-f_{2a}},
\end{eqnarray}
where $w$ is the weighting factor for the normalized objective function $f_1$. 
Note that the objective functions $f_1$ and $f_2$ are normalized with the minimum ($f_{ia}$)  and maximum 
($f_{ib}$) values 
of the functions ($i=1$, 2) for the initial solution population. 
One optimal solution is found for each value of $w$. Generate $S$ solutions, 
$\tilde{\bf X}_j$, $j=1, 2, \cdots, S \ll N$, 
with various weighting factors in order for the solutions to span the Pareto front. 


2. Generate $N$ new trial solutions for the next generation, ${\bf X}^{g+1}_i$, $i=1$, $\cdots$, $N$,  
in the vicinity of the optimal solutions, $\tilde{\bf X}_j$, using a random process, 
e.g., uniformly picking a point within the hypercube centered on $\tilde{\bf X}_j$, with
$|{\bf X}^{g+1}_i -\tilde{\bf X}_j|<\epsilon$. 


3. Evaluate the new solutions with physics simulation to obtain the next data set
${\mathcal D}_{g+1}=\{({\mathbf X}^{g+1}_i,{\mathbf Y}^{g+1}_i)|i=1, 2, \cdots, N\}$. 
Retrain the NN model with all available data set $\{ {\mathcal D}_{0}, {\mathcal D}_{1}, ..., {\mathcal D}_{g+1}\}$ to obtain the new surrogate model 
${\mathbf Y}={\mathcal M}_{g+1}({\mathbf X})$.

4. Stop if ${\mathcal D}_{g+1}$ has converged to the Pareto front. Otherwise 
set $g\leftarrow g+1$ and continue. 

It is critical to exploit the current knowledge of the system to build an accurate surrogate model 
and to find solutions on the Pareto front of the model. To this end, we use all the available solutions 
in the training of the NN model. 
Ensemble modeling is recommended, as it generally performs better than any single model~\cite{Dietterich}. 
Exploration of new parts of the parameter space is achieved by generating the new trial solutions as deviations from the Pareto front. The exploration in the vicinity of the Pareto front 
can mitigate the negative impact of model inaccuracy that results from insufficient sampling around the solution. For example, results in Ref.~\cite{EdelenML} show that the quality of solutions obtained with surrogate models made only from uniform random sampling can suffer substantially when too few samples are chosen, thus necessitating the acquisition of additional samples closer to the optimum to allow further improvement.  This highlights the importance of striking a balance between exploiting previously-learned information  and exploring new parts of the parameter space to efficiently reach an accurate solution (which we address in this work).

We apply the new method to the 
nonlinear beam dynamics optimization of the SPEAR3 storage ring. 
SPEAR3 is a middle energy (3~GeV) third generation light source. Its lattice consists of 
18 double-bend achromat cells, with 14 standard cells and 4 matching cells, each of which 
contains a pair of focusing sextupoles (SF) and a pair of defocusing sextupoles (SD). 
The power supplies of the sextupole magnets are combined into groups according to symmetry, resulting in 
5 SF families and 5 SD families, and a total of 10 sextupole knobs. 
These knobs are needed to optimize the DA and LMA of an emittance upgrade lattice, which reduces the 
horizontal emittance from 10~nm to 7~nm by increasing the horizontal tune by one unit. 
In previous studies ~\cite{HuangPSO}, both MOGA (with the NSGA-II algorithm) and PSO were used to optimize the lattice with 
the sextupole knobs.

\begin{figure}[!hbt]
\centering
\includegraphics[width=4.8in]{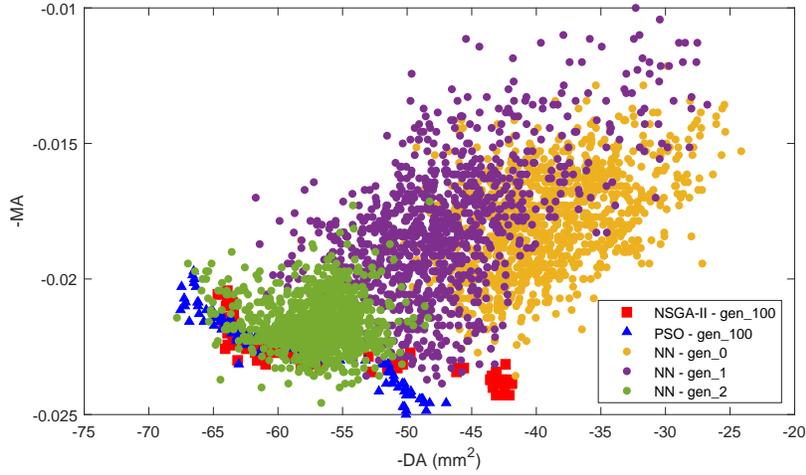}
 \caption{\label{figDAMA1000} Evolution of the two objective functions 
 (dynamic aperture and momentum aperture 
 with negative sign) obtained with NN models for two generations for $N=1000$. The best solutions by MOGA (NSGA-II) and PSO 
 with a population of 100 for 100 generations are shown in comparison. 
 }
\end{figure}

\begin{figure}[!hbt]
\centering
\includegraphics[width=4.8in]{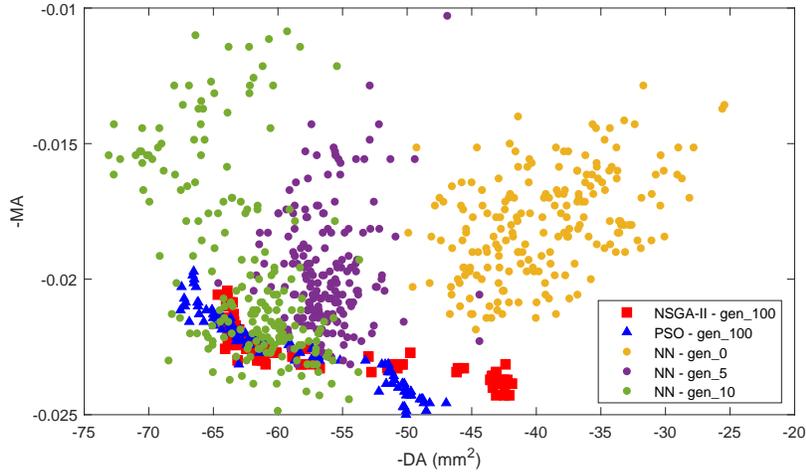}
 \caption{\label{figDAMA200} Evolution of the two objective functions for 10 generations for $N=200$, in comparison to the 
 same MOGA and PSO final solutions. 
 }
\end{figure}

In this study, a slightly different linear lattice is used in the optimization study. 
MOGA and PSO optimizations are used, with 
a population of 100 solutions and 100 generations. 
The final optimal solutions for PSO and MOGA are similar (shown in 
FIG.~\ref{figDAMA1000} and FIG.~\ref{figDAMA200}). 

In the NN approach, a NN model is built with 5 fully-connected, feed-forward layers, 
including the input and output layers. 
To effectively simulate the nonlinearity of the problem, a leaky rectified linear unit activation function  (LeakyReLU)~\cite{LeakyReLU} 
is applied to all intermediate layers. An ensemble of 20 NN models is built, with 
the slope coefficient of the LeakyReLUs randomly picked from within ${(0, 1)}$ for each model. The output of the ensemble model is the average of the 20 NN models.
The ensemble approach increases the reliability of the surrogate model. 
A set of 20 solutions on the Pareto front from the surrogate model 
is obtained using the NSGA-II algorithm, 
with the weighting factor ${w}$ uniformly spanning over the (0, 1) zone. 
The new trial solutions are then generated randomly within the hypercube around these solutions as 
described in step 2 of the algorithm.


To test the method, we initially set the number of solutions in the population to $N=1000$ (generation $g=0$). The objective functions, negated DA and LMA, of the first three generations are plotted in FIG.~\ref{figDAMA1000}, 
in comparison with the final optimal solutions of MOGA and PSO methods. 
The NN solutions in generation $g=2$ are already comparable to the optimal solutions of the two previous methods. 

With a population of $N=1000$, the algorithm converges within two generations, as the surrogate model obtained from 2000 solutions is sufficiently accurate. To reduce the number of total samples used, we repeated this procedure using  $N=200$ and carried out the multi-generation NN method for 10 generations. 
FIG.~\ref{figDAMA200} shows the objective functions for generations $g=0, 5$ and 10. 
The best solutions at generation 10 have exceeded the MOGA and PSO optimal solutions, with only 2000 samples required. In contrast, the MOGA and PSO solutions converged with approximately 10000 samples. 

Since the computation cost of training NN models is negligible for small data sets, the efficiency of the 
optimization method is measured by only the number of solution evaluations. One could start with either a 
large or small population. With a larger population of  solutions, a more accurate model can be trained and the 
algorithm  converges  with fewer generations, although the number of evaluations in 
one generation is higher. With a smaller population size, the model will be less accurate and it generally takes 
more generations to converge. But, it could still achieve 
similar or even better convergence speed in terms of the total number of evaluations, as each generation 
takes fewer evaluations. 
The results shown in FIG.~\ref{figDAMA1000} and \ref{figDAMA200} indicate that for the SPEAR3 nonlinear 
lattice optimization problem, it takes 3000 solutions to find the best solutions for a population size of 
1000, while it takes 11 generations for the $N=200$ case (with a total number of evaluations of 2200). 
For the latter case, the final population extends further into the area with a larger dynamic aperture but also has a poorer momentum aperture. This is probably 
because with fewer initial random solutions (at generation 0), the surrogate model is less accurate 
and thus the new trial solutions are somewhat biased. 




Our study demonstrates that optimization using a 
multi-generation NN surrogate model is applicable to nonlinear storage ring
 design optimization. The NN-based approach enabled optimal solutions to be found with fewer simulation evaluations than PSO and MOGA (e.g. 2000 samples vs 10000 samples). Given the generality of the optimization method, we believe its application can be extended to other design optimization problems, where a global search for the Pareto front in a multi-dimensional parameter space is needed and the evaluation of solutions is computationally costly. 


This  work was supported by the U.S. Department of Energy, Office of
  Science, Office of Basic Energy Sciences, under Contract No.
  DE-AC02-76SF00515 and 
  and DOE contracts 2018-SLAC-100469 and 2018-SLAC-100469ASCR.

\bibliography{da_ref.bib}


\end{document}